\documentclass[preprint,aps]{revtex4}
\usepackage{epsfig}
\usepackage{amssymb}
\begin{document}

\title{Spot foreign exchange market and time series}


\author{Filippo Petroni}
\author{Maurizio Serva}
\affiliation{Dipartimento di Matematica and I.N.F.M. Universit\`a
dell'Aquila, I-67010 L'Aquila, Italy}

\date{\today}

\begin{abstract}
We investigate high frequency price dynamics in foreign exchange market using
data from Reuters information system (the dataset has been provided to us by Olsen
\& Associates). In our analysis we
show that a na\"ive approach to the definition of price
(for example using the spot midprice) may lead to wrong conclusions
on price behavior as
for example the presence of short term covariances for returns.

For this purpose we introduce an algorithm which only uses the non
arbitrage principle to estimate real prices from the spot ones.
The new definition leads to returns which are i.i.d. variables and
therefore are not affected by spurious correlations. Furthermore,
any apparent information (defined by using Shannon entropy)
contained in the data disappears.
\end{abstract}
\maketitle

%
%
%

\section{Introduction}

A foreign exchange market is an over the counter (OTC) market
not subject to any time restriction, in fact, it is open 24
hours a day. Given also that it is the most liquid market in the
world and the availability of tick-by-tick quotes, foreign
exchange market is very convenient for the study of high frequency
behaviors.

Foreign exchange market is made up of about 2000 financial
institutions around the globe which operates by selling or buying
certain amount of a given currency. A market maker (any of the
financial institutions which make the market) is expected to quote
simultaneously for his customers both a bid and a ask price at
which he is willing to sell and buy a standard amount of a given
currency. Each of the major market makers shows a running list of
its main bid and ask rates, and those rates are displayed to all
market participants. In principle each rate from each market
maker is valid until a new rate is displayed by the same market
maker. In practice, this is not the case and no information is
given about the lifetime of each quote.

In analyzing recorded financial data \citep{Tay, Bro+91},
 a difficult and puzzling problem is to
define which is the real asset price. In principle, three different
quotes for the asset are available: bid, ask and traded price (the
price at which the transaction is actually made). Using a wrong
definition for asset price can lead to wrong evaluation of price
dynamics. For example, if the traded price is used to analyze
price dynamics a random zero mean oscillation around the real
price will be found at very short time scale.

We analyze the DEM/USD exchange rates taken from
Reuters' EFX pages (the dataset has been provided to us by Olsen
\& Associates) during a period of one year from January to
December 1998. In this period 1,620,843 quotes entries in the EFX
system were recorded. The dataset provides a continuously updated
sequence of bid and ask exchange rate quotation pairs from
individual institutions whose names and locations are also recorded.
EFX dataset does not contain any information on traded volume and
on the lifetime of quotes. Furthermore EFX quotes are indicative
and they do not imply that any amount of currency has been
actually traded.

The aim of this work is to find the best definition for the asset
price. We start analyzing raw data assuming that the asset price is simply
given by spot quotes. We find that this lead to an indeterminacy of asset price at
very short time scale and to spurious correlations for returns. 
We investigate one possible explanation
assuming that spot quotes contain  an estimation error made by the market maker on the
real price. In this way we do not find the real price but then we introduce an algorithm
which, reducing the spread between bid and ask quotes, is able to
determine the real price and solve the indeterminacy. In the
last chapter we use information theory to strengthen our results.
The key of our work is that we are able to determine the
real price with a parameter free algorithm which uses only
the non-arbitrage principle.

\section{A na\"ive approach to the study of FX microstructure}
The aim of this section is to show that a na\"ive approach to the
analysis of foreign exchange market may lead to wrong conclusions
on price dynamics.

We analyze data taken from EFX Reuters' information system of
DEM/USD exchange quotes of the entire year 1998. In the dataset
each bid ad ask quotes as given by the market operators are
recorded. The dataset does not contain information on trading
prices or on volumes of currencies traded but only tick-by-tick
exchange rates. Prices are irregularly time-spaced and we decided,
 instead of sampling the data in some arbitrarily fixed
sampling time, to use business time as our time flow index 
(see \citep{Ber+03} for an exhaustive
investigation of the problem). According
to our choice $t$ takes all integer values up to $N$ which is the number of quotes
in the dataset.

We indicate with $S^{(a)}_t$ and $S^{(b)}_t$ respectively bid and
ask quotes at time $t$. For our analysis we consider spot
price as given by the average of bid and ask quotes
$S_t=(S^{(a)}_t+S^{(b)}_t)/2$. We stress that this choice for the
spot price is not stringent, the same results can be obtained
if bid or ask quotes are used.

\begin{figure}
\epsfysize=11.0truecm \epsfxsize=11.0truecm
\centerline{\epsffile{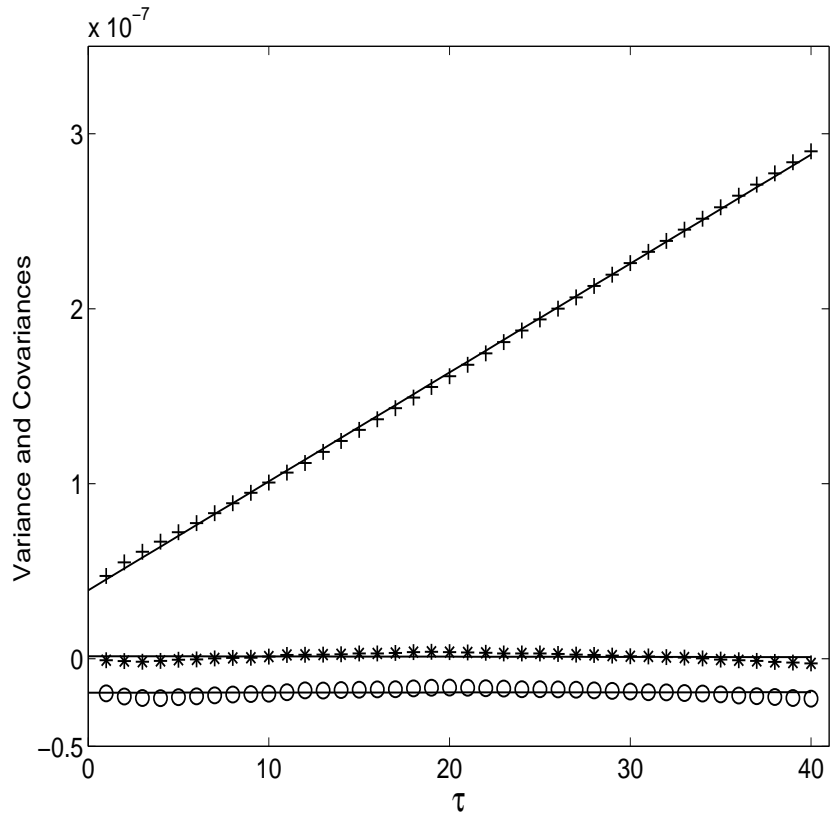}} \caption{DEM/USD spot
exchange rates: variance (\ref{var})(crosses) compared with a
linear fit 2A+B$\tau$, neighboring covariance
(\ref{neiauto})(circles) compared with -A, non-overlapping
covariance (\ref{novaut})(stars) compared with zero.
A and B are identified with $<\epsilon^2_t>$ and $<\tilde{r}^2_t>$.}\label{fig1}
\end{figure}

We define return at two consecutive business time as:
\begin{equation}
r_t\equiv \ln \frac{S_{t+1}}{S_t}
\end{equation}
and, in general, returns at time $t$ and lag $\tau$ as
\begin{equation}
r_t(\tau)\equiv \ln \frac{S_{t+\tau}}{S_t}.
\end{equation}
We estimated using the above cited dataset the $\tau$ dependent
variance of returns:
\begin{equation}
<r^{2}_{t}(\tau)>,\label{var}
\end{equation}
the neighboring covariance of two consecutive returns after
$\tau$ lags
\begin{equation}
<r_{t+\tau}(\tau)r_{t}(\tau)>\label{neiauto}
\end{equation}
and the non-overlapping covariances of returns
\begin{equation}
<r_{t+\tau+s}(\tau)r_t(\tau)>\label{novaut}
\end{equation}
where $s\geq 1$. In the three definitions $<\cdot>$ indicates an
average over the probability distribution. Results are shown in
figure \ref{fig1}. The variance of returns is a linear function of
time lags $\tau$, as expected, but it is different from zero in the limit
$\tau\rightarrow 0$. This imply the existence of an implicit indeterminacy in
the price estimation for vanishing time lags. The same
indeterminacy is responsible for the negative covariance of
two consecutive returns (see below).

In order to explain the previous facts, it has been suggested
\citep{Pas+00} that the spot price is the composition of two
different stochastic processes: a real price change and a noise
contribution which is the result of erroneous evaluations of the
real price by the market operators.

Given that $S_t$ is the spot price at business time $t$ we can
express the two contributions as:
\begin{equation}
S_t=\tilde{S}_t e^{\epsilon_t}
\end{equation}
where $\tilde{S}_t$ is the real price and $\epsilon_t$ is the
error contribution to the real price ($\epsilon_{t}\equiv \ln
(S_{t}/\tilde{S}_t),~\tilde{r}_t=\ln(\tilde{S}_{t+\tau}/\tilde{S}_t)
$). The relation between returns is then given by:
\begin{equation}
r_t=\tilde{r}_t - \epsilon_{t} +\epsilon_{t+1}.
\end{equation}

In this framework we can explain the behavior of the variance and
of the other quantities reported in figure \ref{fig1}. In fact,
with the above definitions, the $\tau$ dependent variance can be
calculated analytically:
\begin{equation}
<r^{2}_{t}(\tau)>=2<\epsilon^{2}_{t}>+<\tilde{r}^{2}_{t}>\tau.\label{lin1}
\end{equation}
where it has been assumed that $\epsilon_t$ and $\tilde{r}_t$ are
uncorrelated i.i.d. random variables. The
neighboring covariance of two consecutive returns after
$\tau$ business time
\begin{equation}
<r_{t+\tau}(\tau)r_t(\tau)>=-<\epsilon^{2}_{t}>\label{lin2}
\end{equation}
and the non-overlapping covariances of returns
\begin{equation}
<r_{t+\tau +s}(\tau)r_t(\tau)>=0
\end{equation}
The above picture corresponds exactly to what one can see in
figure \ref{fig1}. Therefore, it can be estimated the experimental
value for $<\epsilon^{2}_{t}>$ which is $(2.0 \pm 0.2) \times
10^{-8}$ and $<\tilde{r}^{2}_{t}>=(0.64 \pm 0.05)\times 10^{-8}$
for the particular dataset analyzed. We stress that equtions
\ref{lin1} and \ref{lin2} give two independent estimation of the
variance allocated in the error contribution. We find that the two
values, computed from data of figure \ref{fig1}, coincide within
errors.

\begin{figure}
\epsfysize=11.0truecm \epsfxsize=11.0truecm
\centerline{\epsffile{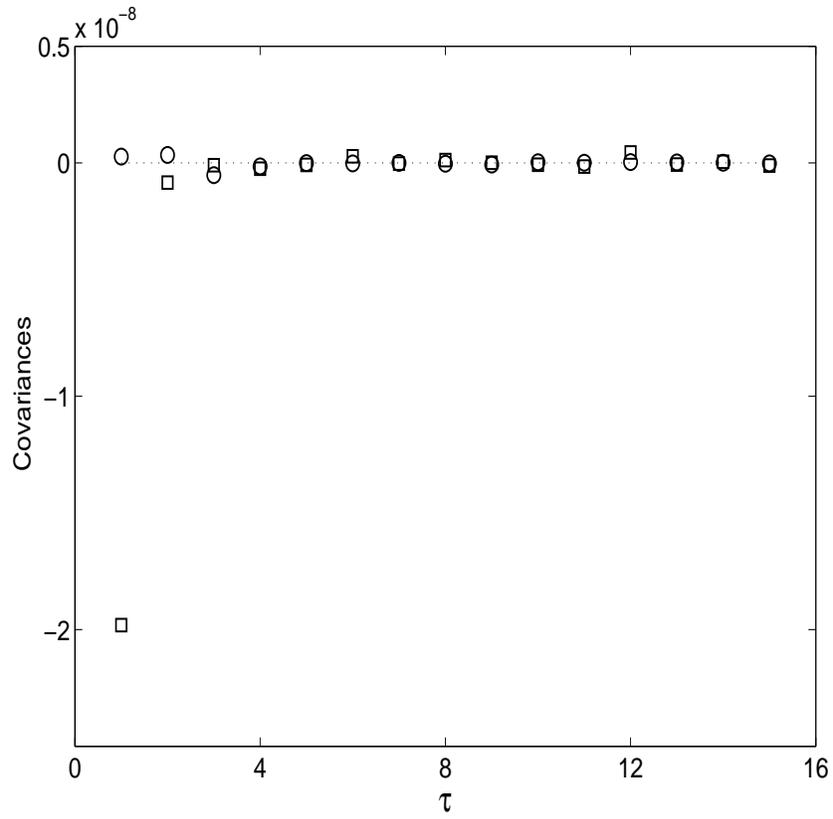}} \caption{Covariance 
$<r_t r_{t+\tau}>$ and $<\tilde{r}_t\tilde{r}_{t+\tau}>$ for
spot (squares) and real (circles) returns.}\label{fig3}
\end{figure}

In order to complete our picture we also estimated the
covariance function on time intervals $\tau$, defined as
\begin{equation}
<r_{t+\tau}r_{t}> \label{autocorr}
\end{equation}
where we considered $<r_{t}>=0$. Results are plotted in figure
\ref{fig3}. The figure shows that the spot returns are one step
negatively correlated ($<r_{t+1}r_{t}>=-<\epsilon^{2}_{t}>$) while
for $\tau>1$ we have $<r_{t+\tau}r_{t}>=0$ according to previous
findings \citep{Jon+98}.

\section{A more realistic approach}
The aim of this work is to find a possible algorithm which is able
to separate the two contributions in the spot price. The
algorithm should be able to solve the indeterminacy found when the
spot price is used to analyze high frequency price dynamics.
From the previous paragraph we have constraints on the variance
allocated in the real price and in the error distribution,
the algorithm should then take this constraints into account.

In DEM/USD 1998 dataset, each quote at each business time is associated with the
financial institution which fixed that quote. In principle this
quote should be valid until the same bank gives a different
exchange rate (both for bid and ask prices). In practice between
two different quotes from the same bank there are several quotes
fixed by other institutions around the world. This suggests that a
bank quote elapses after a certain time even if a new quote has
not been fixed by the same bank. If the dataset contained
information on the time duration of each quote there would be no
problem in establishing real price at each time: it would be
the best bid and ask quotes valid at that time. But this
information is not available and a different strategy has to be
found to establish real price at each time.

The algorithm we propose is the following: let us suppose that we
are observing the bid and ask price of a given currency and that
we are able to detect each quotes from all the financial
institution in the business time $t$.

We define distance between bid and ask as:
$D_t=S^{(b)}_t-S^{(a)}_t$. Notice that for the non-arbitrage
principle this quantity is greater than or equal to zero.
Considering $k$ time lags previous to business time $t$ we
consider the following distance:
\begin{equation}
D_{t,k}=\min_{i\in\{t-k,t\}}S^{(b)}_{i}-
\max_{i\in\{t-k,t\}}S^{(a)}_{i}.
\end{equation}

For each $t$ our algorithm find $\overline{k}$ which gives
$D_{t,\overline{k}}\geq 0$ and $D_{t,\overline{k}+1}<0$. The real
price is then given by
\begin{equation}
\tilde{S}_t=\frac{\left(\min_{i\in\{t-\overline{k},t\}}S^{(b)}_{i}+
\max_{i\in\{t-\overline{k},t\}}S^{(a)}_{i}\right)}{2}.
\end{equation}
 In this
way we can then define a currency quote at each time. Notice that the number
of steps we have to go backwards in time is only given by the
non-arbitrage principle and it is different for every $t$. Once we
have obtained $\tilde{S}_t$ we can define
$\tilde{r}_t(\tau)=\ln(\tilde{S}_{t+\tau}/\tilde{S}_t)$ and compute all
quantities (variances and correlations) already computed for the
na\"ive price definition.

\begin{figure}
\epsfysize=11.0truecm \epsfxsize=11.0truecm
\centerline{\epsffile{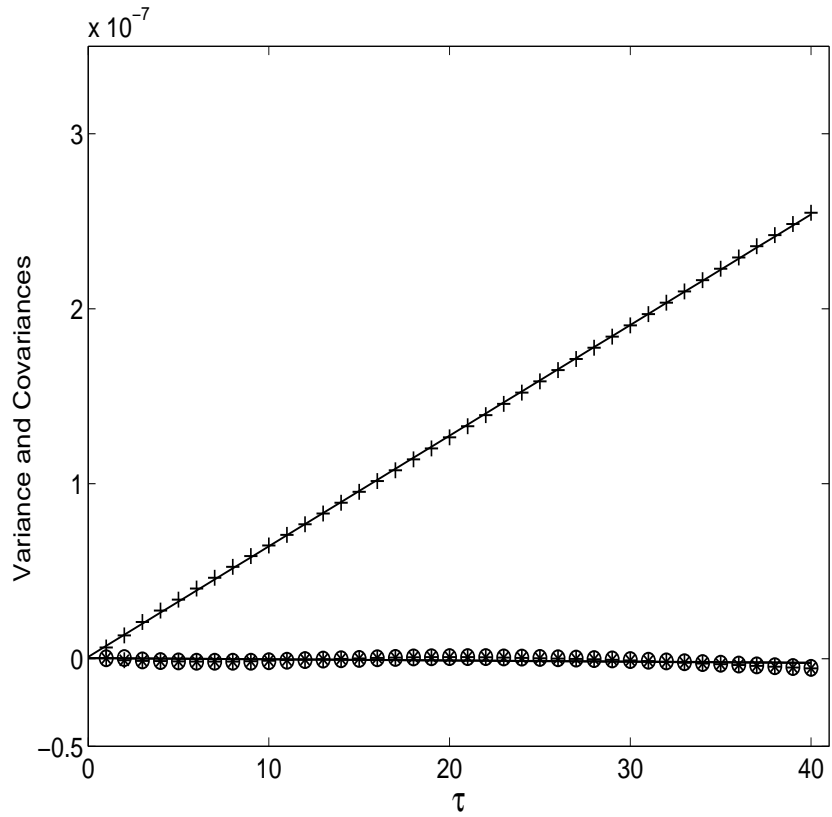}} \caption{DEM/USD real
exchange rates: variance (\ref{var})(crosses) compared with a
linear fit B$\tau$, neighboring covariance
(\ref{neiauto})(circles) compared with zero, non-overlapping
covariance (\ref{novaut})(stars) compared with
zero, B is identified with $<\tilde{r}^2_t>$}\label{fig2}
\end{figure}

As stated above if our algorithm is correct we should have that
the indeterminacy contained in the spot price is removed for the
real price. We then replicate the analysis described in the first
paragraph for the spot price using the above defined real price
$\tilde{S}_t$. Results for this analysis are presented in figure
\ref{fig2}. It can be seen that the variance of returns goes to
zero when business time goes to zero, in fact the experimental
value of $<\epsilon^2>$ in equation \ref{lin1} for the real price is $(3\pm
1)\times 10^{-10}$, two order of magnitude smaller than for the
spot price. Also the neighboring covariance of two
consecutive returns goes to zero. Another interesting results is
that we obtain for the real returns variance a value
($<\tilde{r}^{2}_{t}>=0.64\times 10^{-8}$) which is identical,
within error, to the one predicted in equation \ref{lin1}.

If we estimate the covariance of returns as defined in
equation \ref{autocorr}, we obtain that the real price returns are
uncorrelated at every step (see figure \ref{fig3} where
covariance is compared with that of `na\"ive returns' given
in equation \ref{autocorr}).

The idea we have used here is indeed very simple, we assume that
old quotes are still valid until they produce arbitrage. In spite
of the simplicity we are able to remove all artifacts in the data.

\section{Information Analysis} To be able to perform information
analysis on our dataset first of all we need to code the original
data in a sequence of symbols \citep{Bav+01}. There are several
way to build up such a sequence: one should make sure that this
treatment does not change to much the structure of the process underlying
the evolution of financial data. A partition process of the range
of variability of the data is needed in order to assign a
conventional symbol to each element of the partition. A symbol
corresponds then unambiguously to each element of the partition.
The procedure described below permits to code financial data in a
sequence of binary symbols from which is then possible to quantify
available information.

We fix a resolution value $\Delta$ and define
\begin{equation}
r_{t_i}(\tau)\equiv \ln \frac{S_{t_i+\tau}}{S_{t_i}}
\end{equation}
where $t_i$ is a given business time. We wait until an exit
time $\tau_i$ such as
\begin{equation}
|r_{t_i}(\tau_i)|\geq \Delta
\end{equation}
In this way we only consider market fluctuations of amplitude
$\Delta$. We can build up a sequence of $r_{t_i}(\tau_i)$, where
$t_{1}=t_0+\tau_0$ and $t_{i+1}=t_i+\tau_i$, then we code this
sequence in a binary code according to the following rules:
\begin{equation}
 c_{k} = \left\{
 \begin{array}{lll}
 -1 & if & r_{t_i}(\tau_i)<0 \cr
+1 & if & r_{t_i}(\tau_i)>0 \cr
 \end{array}\right.
\end{equation}
The procedure described above corresponds to a patient investor
who waits to update his investing strategy until a certain
behavior of the market is achieved, for example, a fluctuation of
size $\Delta$.

Once we have build a symbolic sequence we can estimate the entropy
which is defined, for a generic sequence of $n$ symbols, as:
\begin{equation}
H_n=-\sum_{C_{n}}p(C_n)\ln p(C_n)
\end{equation}
where $C_n=\{c_1\dots c_n\}$ is a sequence of $n$ objects and
$p(C_n)$ its probability. The difference
\begin{equation}
h_n\equiv H_{n+1}-H_{n}
\end{equation}
represents the average information needed to specify the symbol
$c_{n+1}$ given the previous knowledge of the sequence $\{c_1\dots
c_n\}$.

The series $h_n$ is monotonically not increasing and for an
ergodic process one has
\begin{equation}
h=\lim_{n\rightarrow \infty } h_n
\end{equation}
where $h$ is the Shannon entropy \citep{Sha48}. It can be shown
that if the stochastic process $\{c_1\dots c_n\}$ is markovian of
order $k$ (i.e. the probability of having $c_n$ at time $n$
depends only on previous $k$ steps $n-1,n-2,\dots,n-k$), then
$h_n=h$ for $n\geq k$. In other cases either $h_n$ goes to zero
for increasing $n$, which means that for $n$ sufficiently large
the $(n+1)$th-symbol is predictable knowing the sequence $C_n$, or
it tends to a positive finite value. The maximum value of $h$ is
$\ln(2)$ for a dichotomic sequence. It occurs if the process has
no memory at all and the $2$ symbols have the same probability.
The difference between $\ln(2)$ and $h$ is intuitively the
quantity  of information we may use to predict the next result of
the phenomenon we observe, i.e. the market behavior.

\begin{figure}
\epsfysize=11.0truecm \epsfxsize=11.0truecm
\centerline{\epsffile{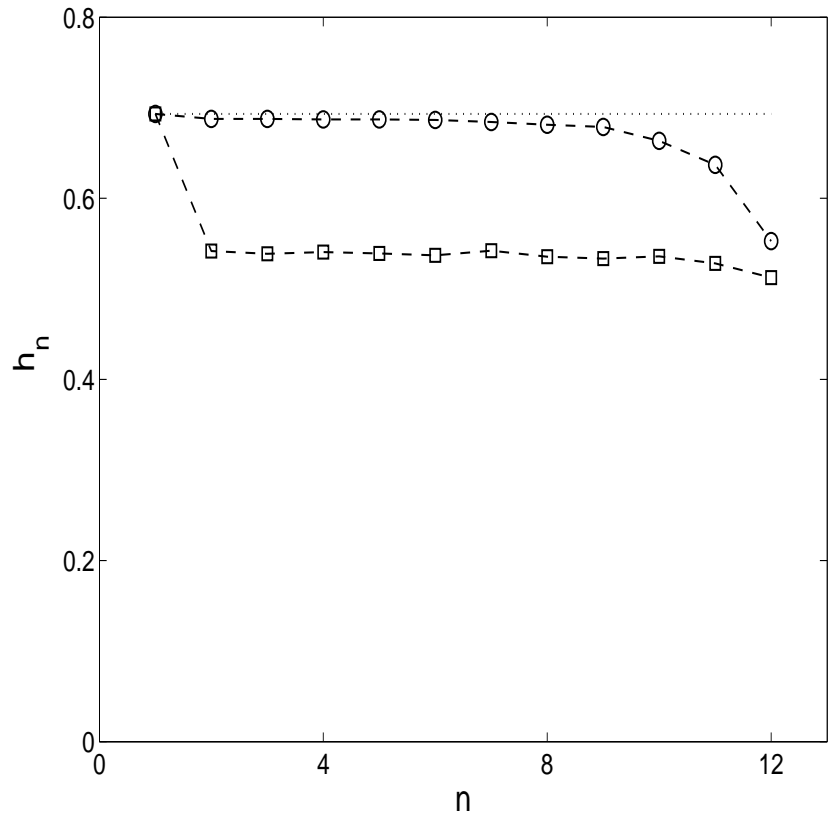}} \caption{Information for spot
(squares) and real (circles) prices}\label{fig4}
\end{figure}

In figure \ref{fig4} $h_n$ is estimated both for real
($\tilde{S_t}$) and spot prices ($S_t$). From the results it
is obvious the different behaviors of the two definition for
currency price. In fact while for the spot price we find a non
zero available information ($\ln 2 - h_n \ne 0$), the stochastic
process is a Markov process of order 1, the real price does
not show this behavior. The available information for the
real price is zero and it remains zero at every step (due to
the finite number of data we can only estimate $h_n$ until
$n\simeq 9$ but we can extrapolate its behavior for $n\rightarrow
\infty$). This show that the real price (unfortunately) is a
stochastic process with no memory and predictability.

\section{Conclusions}
The aim of this work is to find the exact way to extract real prices
from quotes taken form Reuters' Information system. 
Our dataset containes 1,620,843 bid and ask
DEM/USD quotes recorded during the entire year 1998, from the 1st of
January until the 31st of December 1998. 

In section 2  we
show that a wrong behavior of price dynamics can be
obtained when the raw dataset is na\"ively processed. In fact, one
finds an implicit indeterminacy in price specification which
increases the volatility and produces spurious covariances. 
We then explain this
indeterminacy by means of an error contribution which is
responsible for the increased volatility and for the
covariances.

At this point we introduce a parameter free algorithm,
only based on the non arbitrage principle, 
which is able to extract the real prices from the
spot ones. The correctness of the procedure is corroborated by the many results
presented in this work. First of all we show that with the new
price definition the indeterminacy and the one step anti-correlation 
drop to zero. We also show, through
information analysis, that the stochastic process for the new
defined price has no short range memory.

Given our results we think that when studying price dynamics a
strong attention has to be posed on the definition of prices to be
used in the analysis in order to avoid wrong conclusions as, for
example, the existence of short term return correlations.

We stress that we are able to define real prices directly from
spot quotes without the need of further information (time of
validity of quotes) as one could obtain by means of an electronic
broking system \citep{Dan+02}.

In conclusion we would like to propose our method as a general
tool to process raw dataset in order to obtain a new dataset of
the same length whose data are a better representation of price
evolution in the very short time scale.

\section*{Acknowledgment}
We thank Michele Pasquini for illuminating discussions in the
early stage of the present work and for continuous interest and
suggestions.

\newpage

\newpage

\newpage

\newpage
\end{document}